\documentclass[fleqn,usenatbib]{mnras}

\usepackage{newtxtext,newtxmath}

\usepackage[T1]{fontenc}
\usepackage{ae,aecompl}

\usepackage{graphicx}	
\usepackage{amsmath}	
\usepackage{amssymb}	

\newcommand{\magphys}{\textsc{magphys}}

\title[Star formation properties using deep learning]{Predicting star formation properties of galaxies using deep learning}

\author[S. Surana et al.]{
Shraddha Surana,$^{1}$\thanks{E-mail: shraddha.surana@thoughtworks.com (SS)}
Yogesh Wadadekar,$^{2}$, Omkar Bait$^{2}$ and 
Hrushikesh Bhosle$^{3}$
\\
$^{1}$ThoughtWorks Technologies, Yerawada, Pune 411006, India\\
$^{2}$National Centre for Radio Astrophysics, TIFR, Ganeshkhind, Pune 411007, India\\
$^{3}$Centre for Modeling and Simulation, Savitribai Phule Pune University, Ganeshkhind, Pune 411 007, India.
}

\date{Accepted XXX. Received YYY; in original form ZZZ}

\pubyear{2019}

\begin{document}
\label{firstpage}
\pagerange{\pageref{firstpage}--\pageref{lastpage}}
\maketitle

\begin{abstract}

Understanding the star-formation properties of galaxies as a function of cosmic epoch is a critical exercise in studies of galaxy evolution. Traditionally, stellar population synthesis models have been used to obtain best fit parameters that characterise star formation in galaxies. As multiband flux measurements become available for thousands of galaxies, an alternative approach to characterising star formation using machine learning becomes feasible. In this work, we present the use of deep learning techniques to predict three important star formation properties -- stellar mass, star formation rate and dust luminosity. We characterise the performance of our deep learning models through comparisons with outputs from a standard stellar population synthesis code.  
\end{abstract}

\begin{keywords}
galaxies: star formation -- methods: statistical

\end{keywords}



\section{Introduction}

Machine learning algorithms automatically learn from a given dataset without being explicitly programmed. Machine learning refers to any computer program that improves its performance at some task through increased experience (i.e. more example data) \citep{Mitchell:1997:ML:541177}. 
Instead of writing programs to carry out specific tasks, in the supervised machine learning approach, a number of examples of inputs and corresponding outputs are collected and given to the algorithm to learn from. On the other hand, unsupervised machine learning algorithms learn patterns, groupings/ clusters, similarities and/or differences in unlabelled data.

As astronomical datasets grow exponentially in size, machine learning techniques are becoming increasingly useful for creating predictive or classification models which will enable astronomers to expedite the process of astronomical discovery. In recent years, machine learning approaches have been used to tackle a variety of ``big data'' driven problems in astronomy - e.g photometric redshift estimation \citep{Wadadekar2005, Ball2008}, evolution of disk galaxies \citep{Forbes2019}, identification of galaxy morphology \citep[e.g.,][]{Huertas-Company08, Banerji2010, Huertas-Company11}, star-galaxy classification  \citep{Philip2002,Ball2006}, and photometric supernova classification \citep{Lochner2016}, amongst many others. 

Applications of the machine learning approach to prediction of star-formation properties of galaxies have also begun to emerge. \citet{delli2019} estimated the star formation rate for a large subset of Sloan Digital Sky Survey (SDSS) galaxies by training a MLPQNA (Multi Layer Perceptron trained by the Quasi Newton Algorithm) machine learning model. \citet{hemmati2019} use a very different approach by using unsupervised machine learning techniques (Self Organising Maps) to better optimize the model libraries for a given set of observational data. \citet{Stensbo-Smidt2017} estimated specific star formation rates (sSFRs) and redshifts (photo-z's) using only the broad-band photometry from the SDSS. \citet{Baron2019} provides a comprehensive and current overview of the application of machine learning methodologies to astronomical problems.
 
Supervised learning is one form of machine learning where the correct answer (or a truth value) for each data point is known. For each input, its corresponding output is given to the model to train on. The model learns the relationship between the inputs and outputs. The goal is to make the model generic enough that it will be able to accurately predict data points it has not encountered before. In this paper, we have applied deep neural networks which are one of the most widely used supervised machine learning techniques at the present time. 

Deep learning is inspired by the synaptic connections of the brain. Deep learning allows computational models that are composed of multiple  processing layers to learn representations of data with multiple levels of abstraction \citep{lecun2015deep}. These algorithms learn the non linearity in the data for classification or prediction. In this work, we have used deep neural networks to predict three important parameters used to characterise star formation in galaxies viz. stellar mass, star formation rate and dust luminosity.

Deep learning networks generally perform very well in capturing the non linearity in data. Convolutional neural network (CNN), a special type of deep learning network, is now witnessing an explosion of applications in various fields of research, including astronomy. Within the last two years, CNNs have been successfully applied for the classification of galaxy and radio galaxy images in \citet{Barchi2019} and \citet{Lukic2018} respectively. CNNs have also been used in categorization of signals observed in a radio SETI experiment \citep{Harp2019}. \citet{Lovell2019} have also trained CNNs to learn the relationship between synthetic galaxy spectra and high resolution SFHs from the EAGLE \citep{Schaye2015} and Illustris \citep{Vogelsberger2014} simulations.
 
Deep learning has been successfully applied in astronomy to classification problems. \citet{Zhang2006} used Bayesian Belief Networks (BBN), Multilayer Perceptron (MLP) networks and Alternating Decision trees (ADtree) for separating quasars from stars. \citet{Abraham2012} have used a machine learning classifier trained on a subset of spectroscopically confirmed objects, classifying them into stars, galaxies and quasars. \citet{D'Isanto2016} provide a comparison of various machine learning methods for tasks such as  identification of cataclysmic variables, separation between galactic and extragalactic objects and identification of supernovae. \citet{Brescia2015} used MLP with Quasi Newton Algorithm to characterize the physical nature of the large number of objects observed by modern multiband digital surveys. Similarly, deep learning has also been successfully used to build regression models. For example, \citet{Guill2019} have used deep neural networks for the estimation of the muon content of extensive air showers when measured on the ground. \citet{Pearson2019} have implemented deep learning CNNs for estimating strong gravitational lens mass model parameters.

From the diversity of applications mentioned above, it is clear that deep learning approaches have proven useful in tackling a number of astronomical problems. In this work, we attempt to reproduce the capabilities of stellar population synthesis (SPS) models (see \citet{Conroy2013} for a comprehensive review) via a deep learning approach. SPS models combine our understanding of the properties of stars as traced by stellar evolution and star formation histories. The spectral energy distribution of a stellar population in a galaxy is modulated by the presence of  dust which tends to absorb or scatter stellar photons at short wavelengths.  SPS models are physically well motivated and encapsulate a vast and ever growing literature that provides us with ever more sophisticated models. The complexity and sophistication of such modeling comes at a price - finding the best model within the vast parameter space (consisting of millions of possible models) is a compute intensive process. A machine learning approach holds the promise of leaving the building of this complex non-linear model to a machine suitably trained with a large dataset of examples. 
 
In this work, we  train a  deep learning model to predict what are perhaps the three most basic parameters that are used to characterise the star formation process in galaxies  - the current star formation rate (SFR), dust luminosity (DL) and stellar mass (SM). SM is a rough proxy for the integrated star formation in a galaxy throughout its history while the dust luminosity is a measure of the amount of dust in the galaxy.

The data we use  are from the GAMA (Galaxy And Mass Assembly; \citet{Driver2011}) survey. At its core,  GAMA is a spectroscopic survey of $\sim$300,000 galaxies down to $r < 19.8$ mag over $\sim286$ deg$^2$, which was carried out using the AAOmega multi-object spectrograph on the Anglo-Australian Telescope (AAT). The core spectroscopic survey has been greatly augmented by multiwavelength observations of the GAMA fields with a wide variety of telescopes.  We use the GAMA survey Panchromatic Data Release (PDR) catalog \citep{Driver2016} which includes fluxes in upto 21 broadband filters for 120,114 galaxies. For each galaxy the catalog includes position information (RA/Dec), 21 band flux values along with individual errors, a spectroscopic redshift and stellar mass, star formation rate and dust luminosity. 

The last three parameters are determined using a stellar population synthesis code viz.  \magphys{}. This is one of the most widely used SED fitting codes which has been successfully applied on galaxy samples spannig a wide range of galaxy stellar masses, star formation rates and morphological types \citep[e.g.,][]{daCunha10, Berta13, Viaene14, Chang15, Driver2016, Bait17, Driver18}. Our deep learning models are thus limited by the weaknesses of \magphys{} (for instance, the active galactic nuclei's contribution to the SED is not modeled in this code). In regions of parameter space  where \magphys{} is a poor model of the underlying physical reality, our deep learning model will also be a poor representation of the physical reality. Keeping this caveat in mind, we will proceed with our modeling.  For constructing our deep learning model, we used the 21 band data consisting of far ultra-violet (UV) and near UV (from GALEX), the five Sloan Digital Sky Survey bands: $ugriz$, near infrared bands: X, Y, J, H, K, WISE mid-infrared bands: W1, W2, W3, W4, and the five Herschel bands: P100, P160, S250, S350, S500. The spectroscopic redshift listed in the PDR catalog is used as an additional input parameter for our model.
In Section~\ref{methodology}, we introduce the deep learning algorithm and the data preprocessing we have carried out. In Section~\ref{rnd}, we describe and discuss our results and we conclude in Section~\ref{conclusion}.

\section{Methodology} \label{methodology}
\begin{figure}
    \includegraphics[width=\columnwidth]{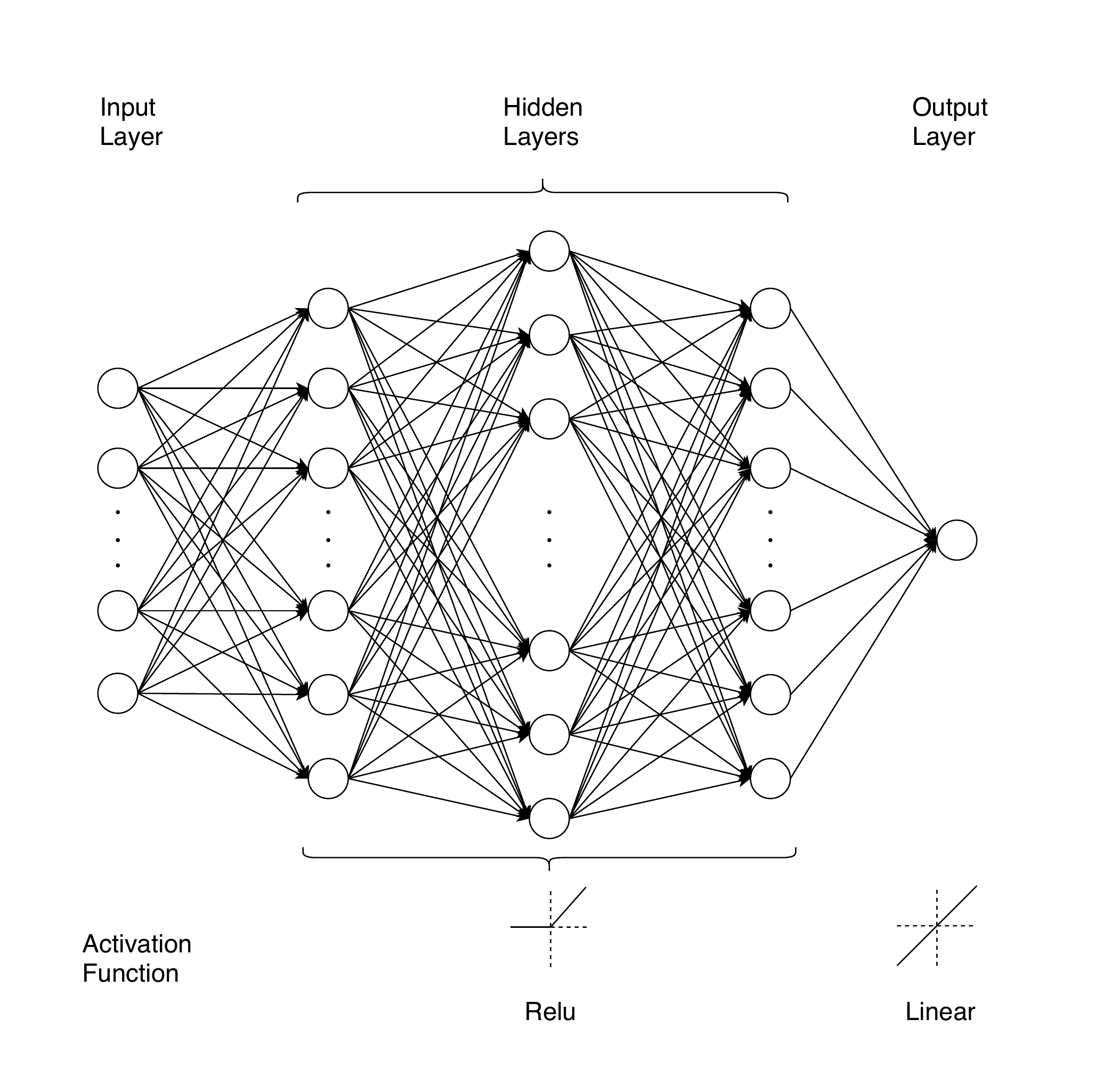}
    \caption{Deep learning architecture used in our work. There are 22 input nodes corresponding to the 21 bands and the spectroscopic redshift. The number of hidden layers and the number of nodes in each layer depend on the individual model and is specified in Table~\ref{tab:architecture_table}.  The output layer will have one node corresponding to the free parameter being predicted. The activation function used is ReLU for the hidden layers. For the output layer, the linear activation function is used.}
    \label{fig:dl_architecture}
\end{figure}

The structure of a typical deep learning architecture is shown in Fig.~\ref{fig:dl_architecture}. The model consists of input nodes - one per input attribute (broadband fluxes and redshift, for our specific problem). For our problem, we predict each free parameter separately and hence there is one output node for each of the three models. There are several hidden layers between these input and output layers. The number of nodes in each hidden layer are typically greater than the number of input nodes. As the model encounters more and more data, it learns patterns in the dataset that distinguishes one outcome from  another, thereby enriching it over time. 

In this section we first describe our sample selection which aims to achieve a good trade off between coverage and homogeneity of the data to be given to the ML model. This is followed by a brief discussion of the current \magphys{} SED fitting technique, and how the various hyper parameters of the deep learning algorithms were derived.

\subsection{Sample Selection}
\begin{figure}
    \includegraphics[width=\columnwidth]{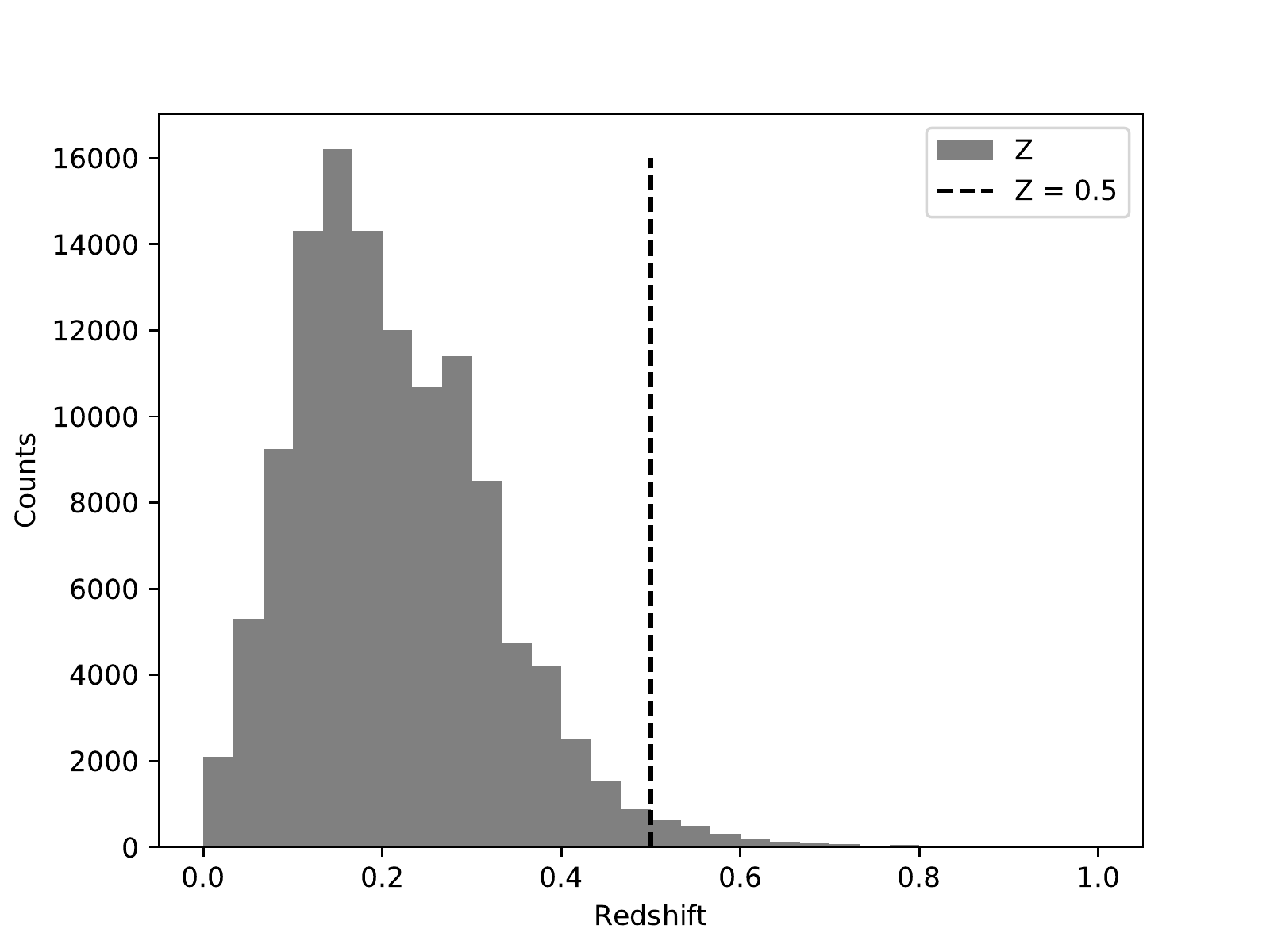}
    \caption{The redshift distribution of galaxies in the GAMA catalog. We only used galaxies with $z < 0.5$ in our analysis which forms more than 98\% of galaxies in the GAMA catalog.}
    \label{fig:redshift_distribution}
\end{figure}

\begin{figure}
    \includegraphics[width=\columnwidth]{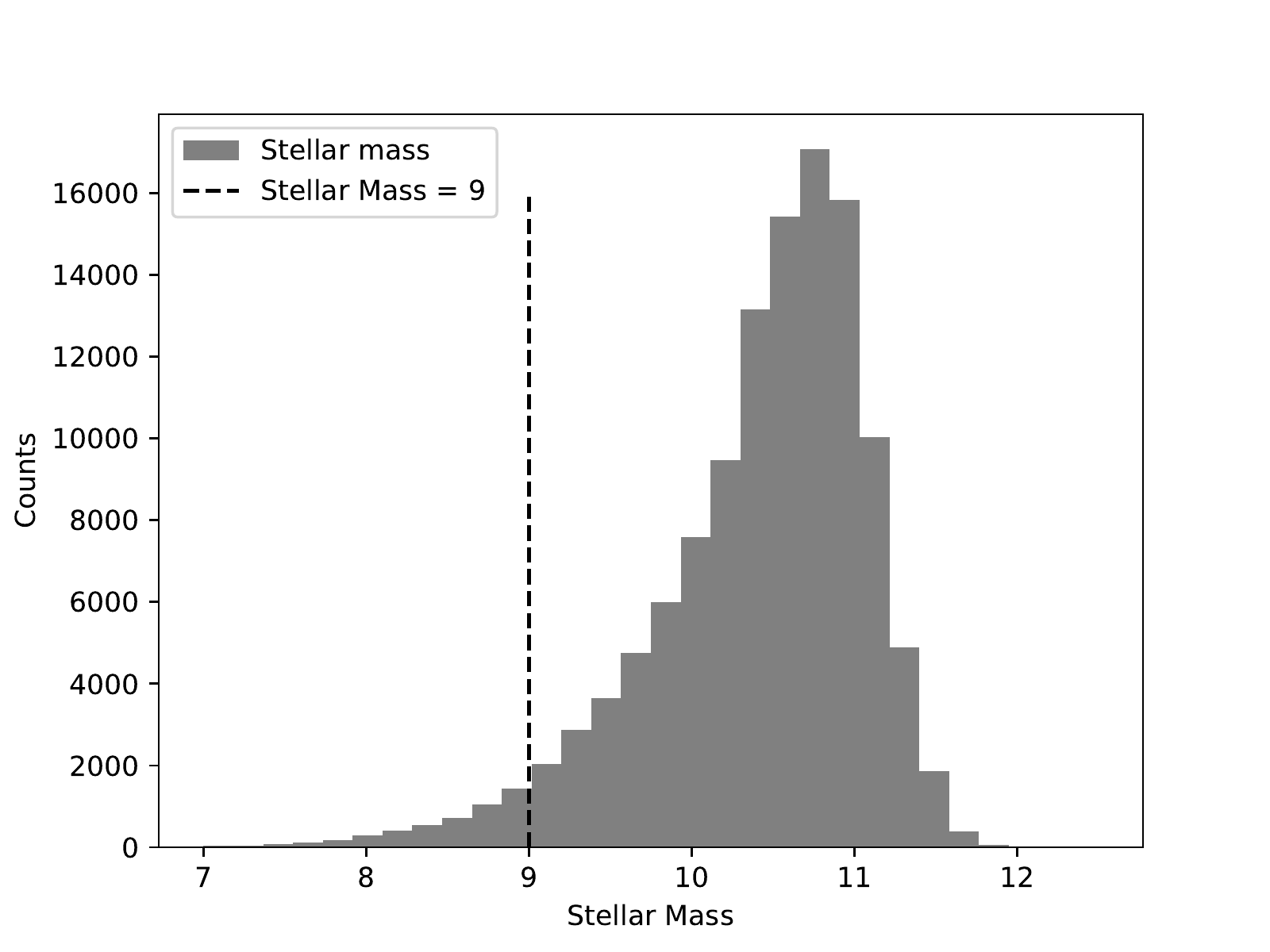}
    \caption{Stellar mass distribution of galaxies in the GAMA catalog. We only used galaxies with stellar mass >  $10^9 M_{\odot}$ which forms more than 95.9\% of galaxies in the GAMA catalog.}
    \label{fig:stellar_mass_distribution}
\end{figure}

We selected all galaxies with  $0<z\leq0.5$ and stellar mass $\geq 10^9 M_{\odot}$ from the GAMA catalog. The distribution of both these parameters - redshift and stellar mass are shown in Fig.~\ref{fig:redshift_distribution} and Fig.~\ref{fig:stellar_mass_distribution} respectively. Only 2,173 galaxies had redshift greater than 0.5 and 4,867 galaxies had stellar mass less that $10^9 M_{\odot}$. By imposing these arbitary redshift and stellar mass cuts, we avoid sparsely populated regions of the parameter space where the small number of training examples makes it difficult to effectively train a deep learning network. We also removed galaxies with missing flux values in any of the 21 filters. There were 36,768 such galaxies that had at least one missing value. We observed a small number (110) of galaxies with negative flux values in at least one filter. We removed galaxies with such unphysical flux measurements, from our modeling.

\begin{figure}
    \includegraphics[width=\columnwidth]{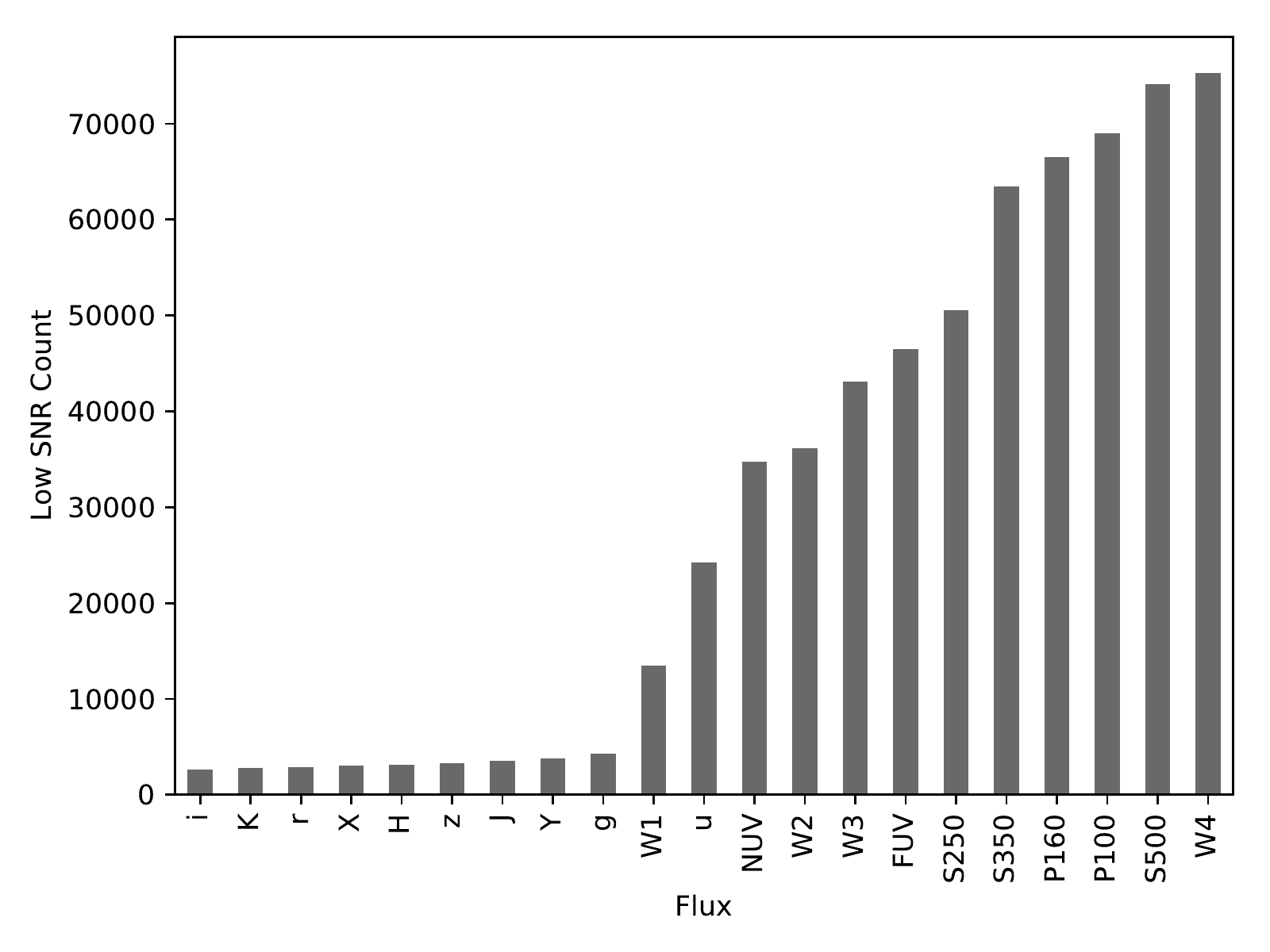}
    \caption{Bar plot showing low SNR counts (SNR < 3) indicating the quality of the data. For most galaxies, SNR is relatively good in the SDSS and near infrared bands and is the poorest in the WISE W4 band and in the far infrared Herschel bands.}
    \label{fig:low_snr_count}
\end{figure}

Inspite of these cuts, there remained a few galaxies with very low signal-to-noise(SNR=$\frac{flux}{flux error}$) values in several bands. Fig.~\ref{fig:low_snr_count} shows the number of galaxies that have SNR less than 3 in each of the input flux bands. Retaining such  unreliable measurements poses problems in the training. After some experimentation, we found that galaxies with at least 6 flux measurements having a SNR of 3 or more gave a good trade off, between obtaining good predictions while only reducing the sample size by a small fraction.  

Post all the filtering described above, we are left with 76,455 galaxies which form our final sample. Note that a galaxy is excluded from our sample only because 1. it occupies a sparsely populated region of the parameter space or 2. it has flux measurements that are missing or unreliable. With future, deeper multiband imaging surveys both of these defects will be adequately addressed.

\begin{figure}
    \includegraphics[width=\columnwidth]{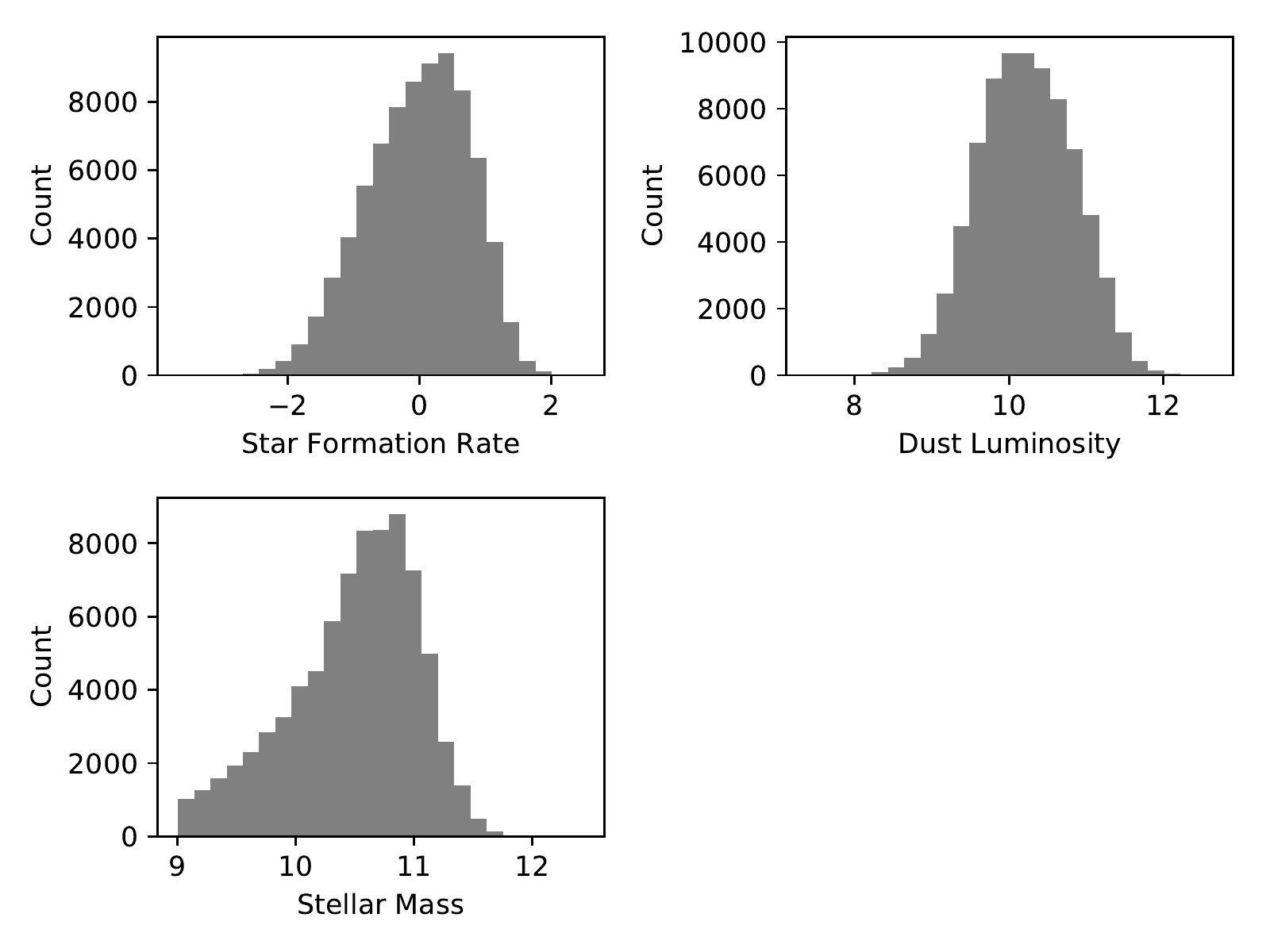}
    \caption{Distribution of the output parameters which we predict using a deep learning model.}
    \label{fig:free_parameter_distribution}
\end{figure}

\subsection{\textsc{MAGPHYS} SED Fitting}
The three output parameters used in this work, stellar mass, star formation rates and dust luminosity are from \citet{Driver2016} which are derived using a stellar population synthesis code \magphys{} \citep{daCunha2008}. \magphys{} estimates various physical properties of a galaxy by fitting its observed SED in ultra-violet (UV), optical, and  infrared (IR) bands with a large library of template SEDs. We briefly describe how \magphys{} creates the library of template SEDs here. For the stellar population synthesis modelling \magphys{}
uses the CB07 library, which is the unpublished version of the \citet{Bruzual2003} model with a Chabrier initial mass function. A set of stellar library are then constructed using a exponentially decaying star formation history with random bursts of star formation superimposed on it. The metallicity is varied from 0.02 to 2 times solar metallicity. This stellar library also undergoes a dust attenuation using the two-component \citet{Charlot2000} model. The amount of energy attenuated due to dust is then re-distributed in the mid to far-IR in various components (in the form of polyaromatic hydrocarbons, hot, warm and cold dust) assuming energy balance. The dust emission is simply modelled as due to a grey body (i.e. modified blackbody) with varying temperatures and emissivities. \magphys{} then performs a $\chi^2$ fit on the observed data with the entire library of template SEDs. The code then builds a probability distribution function for each of the free parameters in the model by weighting the parameter value with the probability given by exp$(-\chi^2_{i}/2)$ for the $i$th model. In our work, we have used the median values (i.e. the 50th percentile of the probability density function) for the several derived physical properties viz. stellar mass, star formation rate and dust luminosity. Fig.~\ref{fig:free_parameter_distribution} shows the distribution of these physical properties which are the output parameters we model in this paper.

\subsection{Deep learning} \label{deeplearning}

We implemented various machine learning models on our dataset including random forest, support vector machine and deep neural networks. After considerable effort in optimising the results with each machine learning model, we found that deep neural networks gave the best results. The remainder of this paper will therefore focus on the results of our models trained using deep neural networks. We explored 2 approaches for predicting the 3 free parameters of star formation history.  The three output parameters having some correlation amongst them, we trained a deep learning model that would predict all the three free parameters together. We also trained a separate model for each of the free parameters. The predictions were more accurate when separate models were created. This may be because, although the free parameters may have some correlation amongst them, they may also have significant differences that one model is not able to capture.
There may not be a single minima in the loss function space for all the three free parameters. Training a separate model for each free parameter allows the optimizer to search for the global minima pertaining to each. Hence, we are able to generate optimised models catering to each individual free parameter.

In this section, we describe the deep neural network model that we have used which is shown in Fig.~\ref{fig:dl_architecture}. The model was implemented using keras \footnote{Chollet F., et al., 2015, Keras, https://keras.io} which gives high level APIs for neural network. It is open source and is written in Python. Implementing a deep learning algorithm however, involves tuning of many hyperparameters. These are:

\begin{enumerate}
    \item Number of hidden layers - This is where the deep of deep learning is. While one or two layers may suffice for simple datasets, complex datasets require more hidden layers. \citep{hinton2006fast}. The relation of the input attributes in the dataset with the free parameters we are trying to predict is not linear. Hence, in all the three models (to predict SM, SFR and DL), three or more hidden layers have worked well.
    
    \item Number of nodes in each hidden layer - Using too few nodes in the hidden layers may lead to underfitting because there are too few nodes to detect the patterns or features in a complicated dataset. 
    On the other hand, using too many nodes may lead to overfitting as the model will learn details and noise pertaining to the training data very well, but will not be able to generalise well to unseen data. Hence, the number of nodes in each hidden layer needs to be derived with some trial and error to make sure the model does not overfit nor underfit the dataset. The number of nodes in each hidden layer for the three parameters are listed in Table~\ref{tab:architecture_table}.
    
    \item Kernel Initializer - This defines the way to set the initial random weights between the layers and can affect the speed of convergence on the optimization algorithm. We have used the 'RandomNormal' initializer which generates random weights with a normal distribution. Training algorithms for deep learning models are iterative in nature and require some initial point from which to begin the iterations. Moreover, training deep models is a sufficiently difficult task and algorithms are strongly affected by the choice of initialization \citep{Goodfellow2016}. These initial weights serve as the starting point to search for the optimal solution and are initialised randomly as the nature of the algorithm itself is stochastic.
    
    \item Activation function - Activation function is a property of a node. It does a (in most cases non-linear) transformation on an input or a set of inputs coming in from the previous layer. The transformed output is given to the nodes in the next layer which again go through the specified activation function for that layer. It acts as a switch for the node to capture the non linear information in the input data that enables it to predict the output. The activation function at the hidden nodes used is ReLu \citep{lecun2015deep, Nair2010RectifiedLU}. \citet{zeiler2013rectified} and \citet{glorot2011deep} state the advantages of ReLU such as easier optimization, fast convergence, better generalization and faster compute time. \citet{sutskever2012imagenet} show an increase in speed of up to 6 times using ReLU in a four-layer CNN. For our regression problem, we have used the linear activation function at the output node. 
    
    \item Loss function - This is also known as cost function or error function. It is the error between the actual and predicted values. The model is trained to minimize this error value. We have minimised the mean absolute error for all the three models as it is more robust to outliers. It is the average of the absolute difference of the actual and predicted values.
    
    \item Optimizer - In order to minimize the loss function, an optimization algorithm is used. The optimizer attempts to find the global minima for a convex loss function. We have used the Adam optimizer which is a first order gradient based optimization of stochastic objective functions, based on adaptive estimates of lower-order moments \citep{kingma2014adam}. Other optimizers were also tried viz. SGD (Stochastic Gradient Descent), RMSprop, Adagrad, Adadelta and others that are built-in in keras. We used the Adam optimizer in our work as it gave best results. 
    
    \item Epochs -  This is the number of iterations to train the model. A smaller value will mean not training the model enough and may lead to an underfitted model. On the other hand, training the model for many epochs may lead to an overfitted model. The number of epochs each model is trained on is different and is mentioned in Table~\ref{tab:architecture_table}
    
    \item Batch size -  The batch size can be anything between 1 and the size of the training sample. Taking a batch size of 1 leads to some wandering around as the gradient of the sample may be in the wrong direction. But on average it will head towards the minima as with a full batch gradient descent, only with greater number of iterations/epocs. The cost of computing in this case is very trivial. A full batch gradient descent, takes into account all the samples at once and computes the best direction towards the minima. It requires more computational power, but gets to the minima in the fewest steps. However, when the data size is large, it cannot be given to the algorithm at once as the computer RAM will not have the capacity to hold all the data. In such cases, the training sample is divided into batches and given to the algorithms, post which the internal parameters of the model are updated. We ran our algorithm with various batch sizes and found 256 to be a good tradeoff between the time taken, accuracy and computational power required.

\end{enumerate}

A separate model was trained for predicting each of the 3 free parameters of star formation history. The deep learning network architecture is different for each of these free parameters. Table~\ref{tab:architecture_table} shows the optimal architecture used to predict each of the free parameters separately. Table~\ref{tab:hyperparameters} shows the values of the common hyperparameters used in the models.

\subsection{Input Data Transformation}
\begin{figure}
    \includegraphics[width=\columnwidth]{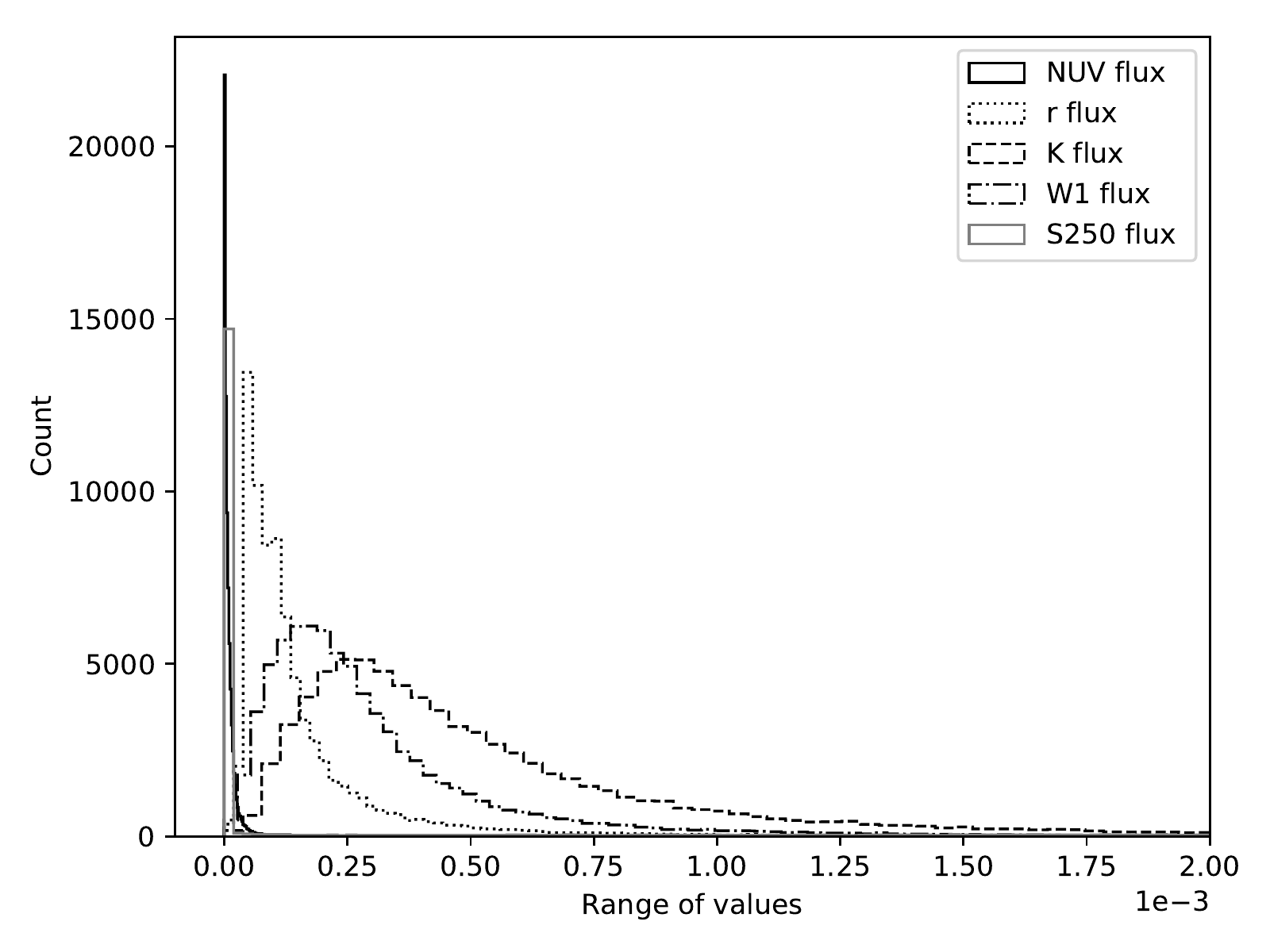}
    \caption{Distribution of a sample of input attributes. The sampled attributes have different range of values. This can lead to attributes with higher numeric values to have greater influence on the result. Fig.~\ref{fig:hist_scaled} shows the transformed values for these same flux bands. Note that the entire distribution of fluxes spans a larger range than shown here.}
    \label{fig:hist_original}
\end{figure}

\begin{figure}
    \includegraphics[width=\columnwidth]{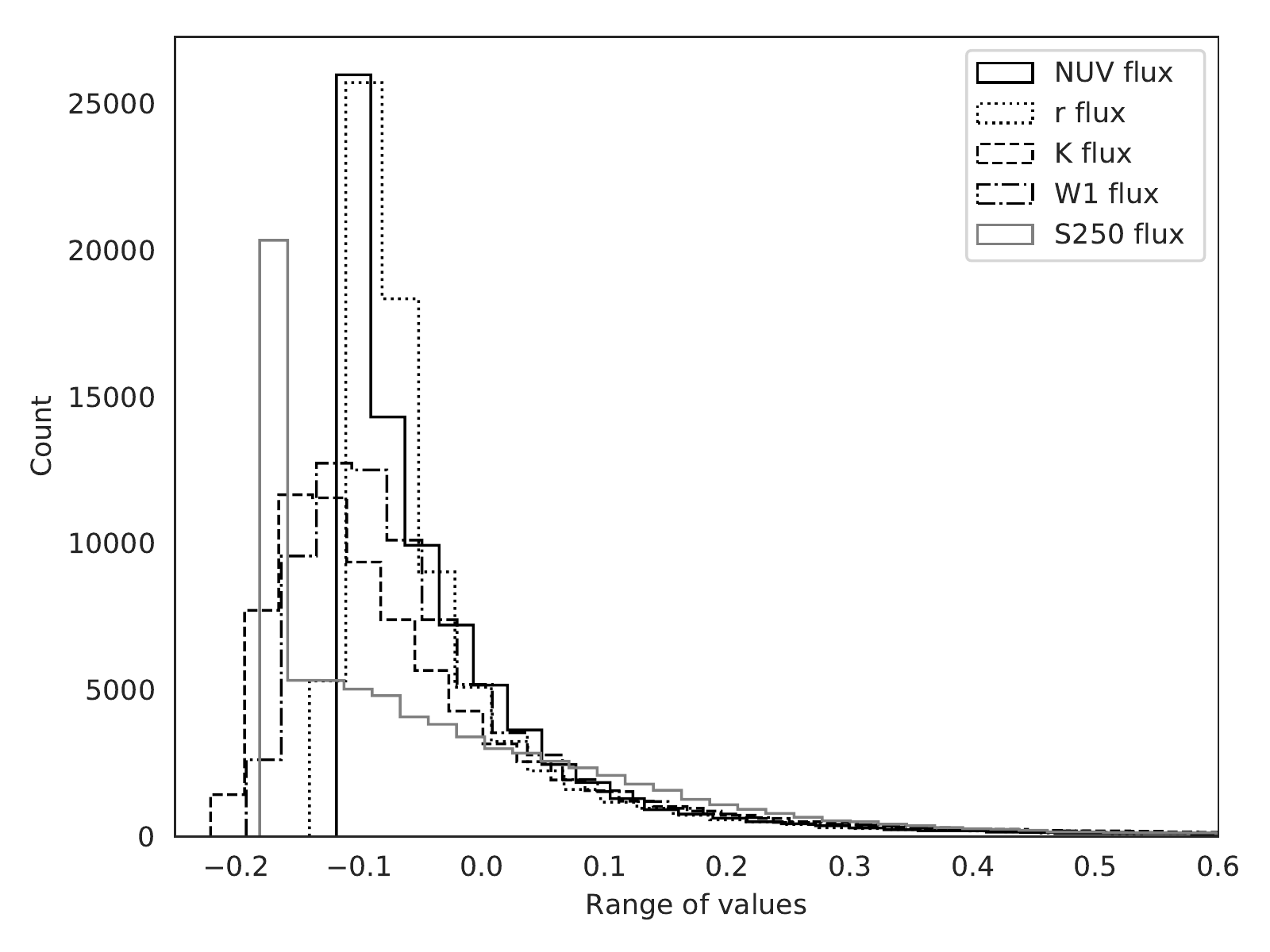}
    \caption{Distribution of a sample of input attributes after transformation operation. All the attributes now have similar range of values with a mean of zero and standard deviation of one. This lets the algorithm see all the attributes with the same level of importance. If transformation is not done in this manner, then attributes that have a high numeric value end up having a higher influence on the model. Note that the entire distribution of fluxes spans a larger range than shown here.}
    \label{fig:hist_scaled}
\end{figure}

It is a common requirement for many machine learning algorithms to transform the input attributes in order to standardise them \citep{witten2016data}. Machine learning algorithms work well on a normally distributed dataset. The attributes in the dataset are standardised by removing the mean and scaling to unit variance. The distribution of some input parameters in the dataset before and after transformation is as shown in Figs.~\ref{fig:hist_original} and~\ref{fig:hist_scaled}

For our experiments we split the data 64\% - 16\% - 20\% into training, validation and test sets. The model is trained on the training set. This model is then validated on the validation set i.e. its error is calculated on the validation set and this error information is used to improve the model. We want the model to generalise well and not overfit the training data. Once the model is finalised, it is given the test data to see how well the model performs on data it has never seen before. This gives a good idea of how well the model will perform in a real life scenario. The performance on this test set is presented in Table~\ref{tab:results_table}

Scikit-learn \citep{Pedregosa:2011:SML:1953048.2078195} which consists of efficient tools for machine learning, statistics and data processing; numpy and pandas \citep{mckinney-proc-scipy-2010} are used for the data transformations and manipulations. The Python version used is 3.7.3.

\begin{table*}
	\centering
	\caption{Hyperparameter values for the deep neural network architecture for each model}
	\label{tab:architecture_table}
	\begin{tabular}{lccc} 
		\hline
		Parameter & Stellar Mass & Star Formation Rate & Dust Luminosity \\
		\hline
		Number of hidden layers & 3 & 3 & 5\\
		Number of nodes in each layer & 22-44-66-66-1 & 22-110-220-110-1 & 22-66-110-220-110-66-1\\
		Epoch & 5000 & 1500 & 1700\\
		Early Stopping Patience & 300 & 500 & 300\\
		\hline
	\end{tabular}
\end{table*}

\section{Results and Discussion} \label{rnd}
Here we present our model performance and its loss i.e. error as the algorithm iterates. We compare the time taken by the deep learning algorithm to that taken by the  \magphys{} modelling. In order to compare the model performance, the error metric used is the standard deviation of the difference between the actual (\magphys{} best fit value) and predicted by our deep learning model: $$error = \sigma(y_{actual}-y_{predicted})$$ Note that all output parameters are computed in logarithmic space. The error and time taken  are given in Table~\ref{tab:results_table}. For completeness we have also reported the values in rmse - root mean square error and mean absolute deviation - mad as well. The $r^2 adjusted$ value is the amount of variance explained by the model.


\begin{table}
    \centering
    \caption{Common hyperparameters used in all the three deep learning models}
    \label{tab:hyperparameters}
    \begin{tabular}{lc} 
        \hline
        Hyperparameter & Value\\
        \hline
        Learning Rate & 0.001\\
        Batch size & 256\\
        Optimizer & adam\\
        Loss function & mean absolute error\\
        \hline
    \end{tabular}
\end{table}

\begin{table*}
    \centering
    \caption{Error and time taken to train for the three models created. The models were trained on a 2.2 GHz Intel Core i7 machine with 32 GB 2400 MHz DDR4 RAM}
    \label{tab:results_table}
    \begin{tabular}{lccc} 
        \hline
        Parameter & SM & SFR & DL\\
        \hline
        $error = \sigma(y_{actual}-y_{predicted})$ & 0.0577 & 0.1643 & 0.1143\\
        root mean square error & 0.0582 & 0.1658 & 0.1159\\
        mean absolute deviation & 0.0399 & 0.0987 & 0.0750\\
        $r^2$ adjusted & 0.9885 & 0.9550 & 0.9644\\
        Time taken to train(seconds) & 1110 & 183 & 477\\
        \hline
    \end{tabular}
\end{table*}

\begin{figure}
    \includegraphics[width=\columnwidth]{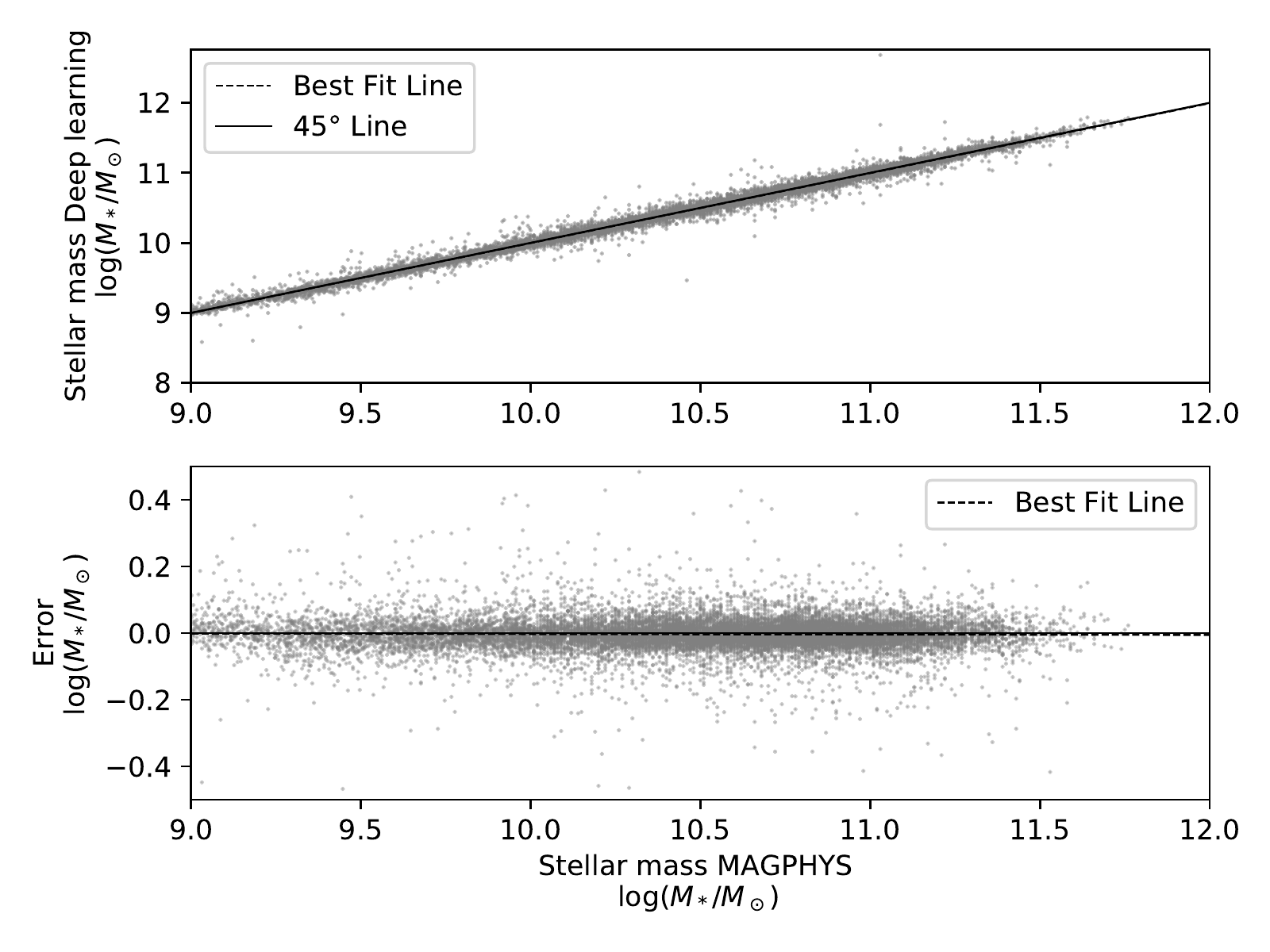}
    \caption{The top scatter plot shows the actual values calculated by \magphys{} compared to the values predicted by the deep learning model for stellar mass. The plot shows the best fit line through the scatter plot and the 45 degree line which would have been obtained if the deep learning model predicted exactly the same values as the \magphys{} model.
    The bottom plot highlights the error across the values of stellar mass. The scatter of points is uniform across the values of stellar mass.}
    \label{fig:sm_fig}
\end{figure}

\begin{figure}
    \includegraphics[width=\columnwidth]{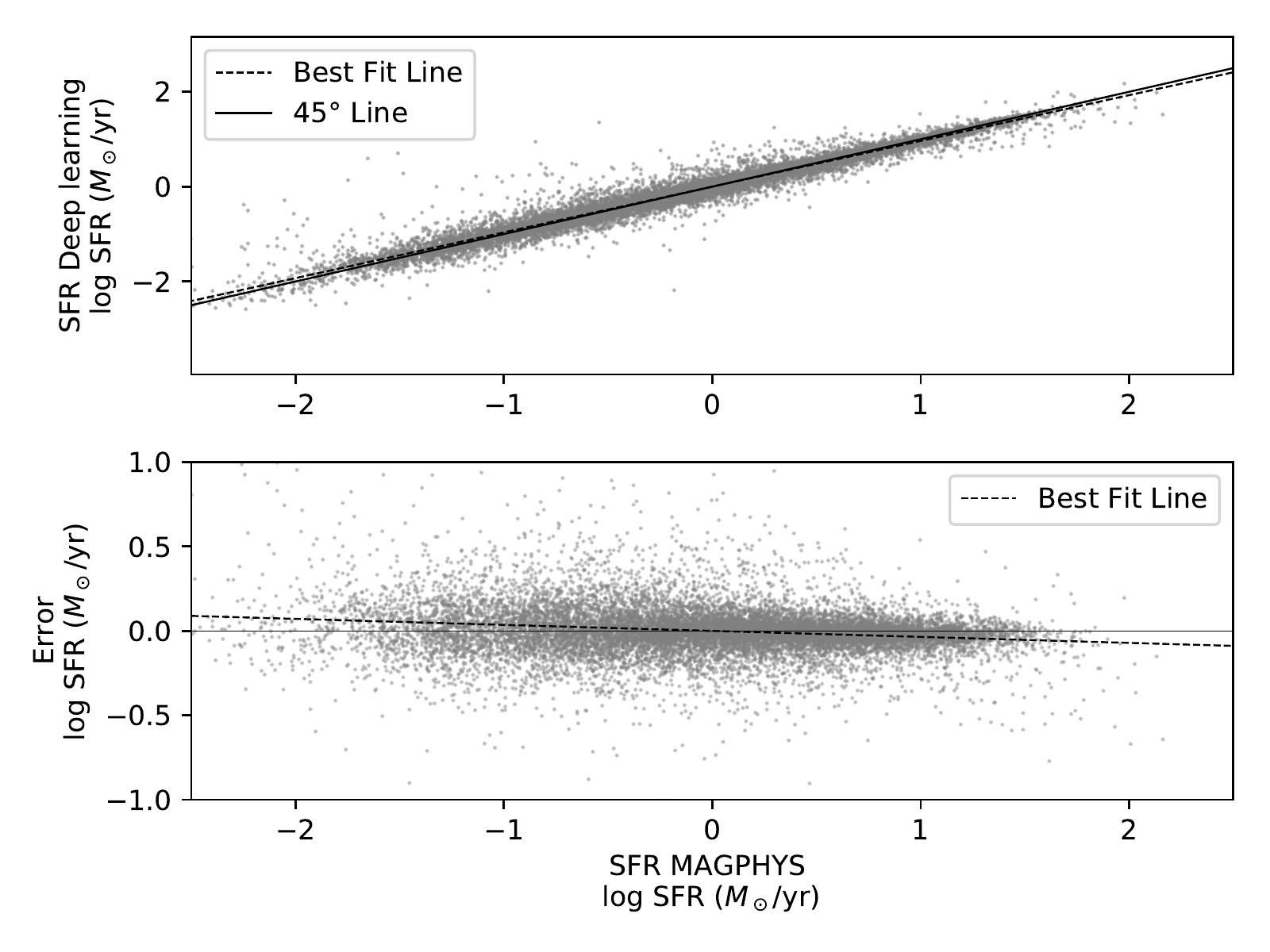}
    \caption{The top scatter plot shows the actual values calculated by \magphys{} compared to the values predicted by the deep learning model for SFR. The plot shows the best fit line through the scatter plot and the 45 degree line which would have been obtained if the deep learning model predicted exactly the same values as the \magphys{} model.
    The bottom plot highlights the error across the values of SFR. It can be observed that there is a higher scatter for lower values of SFR.}
    \label{fig:sfr_fig}
\end{figure}

\begin{figure}
    \includegraphics[width=\columnwidth]{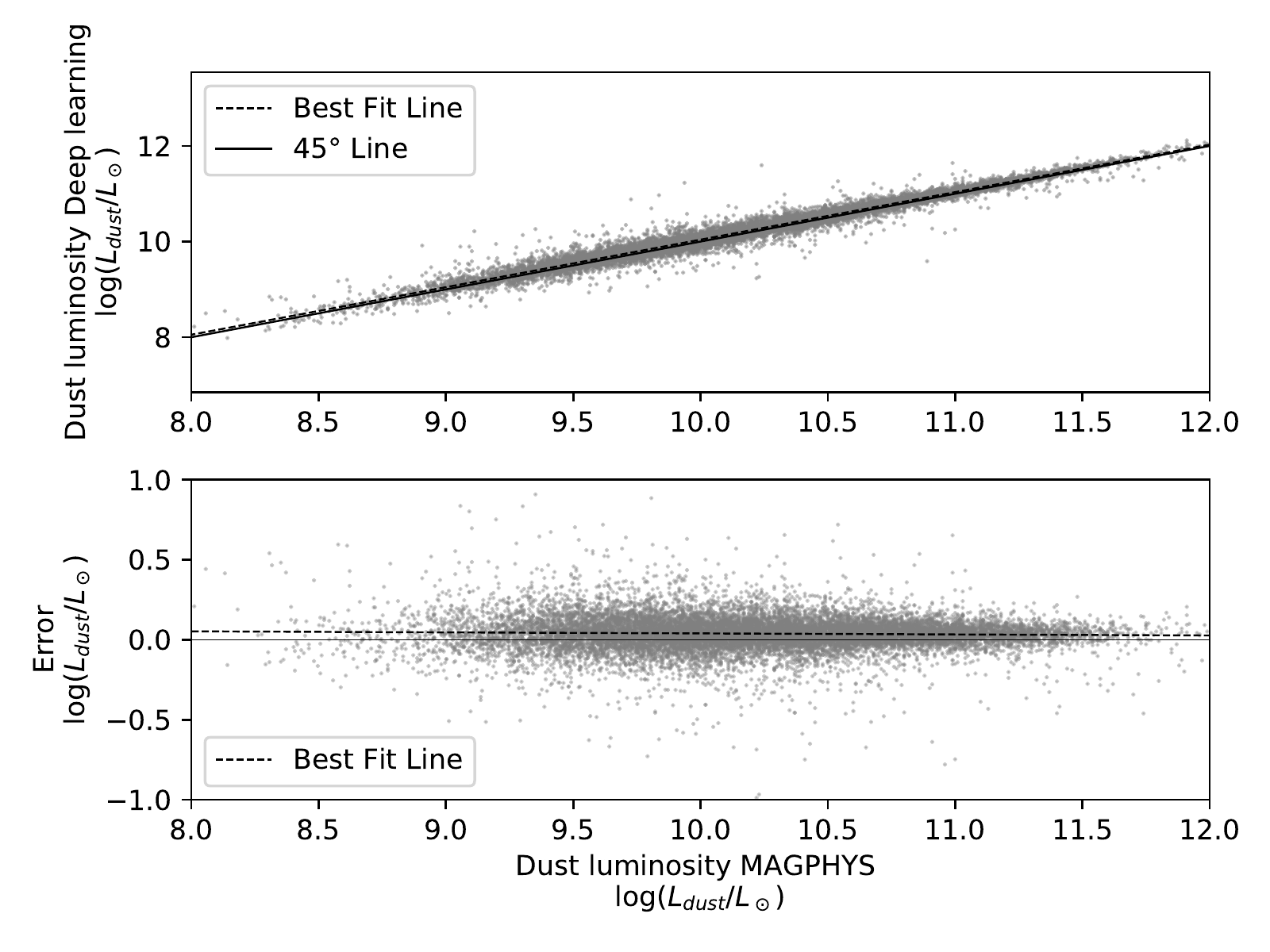}
    \caption{The top scatter plot shows the actual values calculated by \magphys{} compared to the values predicted by the deep learning model for dust luminosity. The plot shows the best fit line through the scatter plot and the 45 degree line which would have been obtained if the deep learning model predicted exactly the same values as the \magphys{} model.
    The bottom plot highlights the error across the values of dust luminosity. It can be observed that there is a higher scatter for lower values of dust luminosity. The errors are also skewed towards the higher side indicating that the model is predicting higher values for lower values of dust luminosity.}
    \label{fig:dl_fig}
\end{figure}

Figs.~\ref{fig:sm_fig}, \ref{fig:sfr_fig} and  \ref{fig:dl_fig} show the scatter plot (top) and the difference plot (bottom) between the predicted and actual value for stellar mass, SFR and dust luminosity respectively. The scatter plot shows the best-fit line (the dashed line) between the predicted and actual values. The plot also shows the 45 degree line - the line when predicted values are exactly equal to the actual values. As can be seen, the fit for stellar mass is very close to the 45 degree line with very low scatter. Similar is the case for dust luminosity. Star formation rate has a little more scatter compared to the former two free parameters. The difference plot highlights this. It is important to note that the values of these free parameters span three orders of magnitude. The same model is able to predict over this entire range of values as deep learning models are able to capture the non linear relationships between the input flux values and the output free parameters. The scatter in the difference plot of SFR is more for lower values of SFR. This systematic trend needs further investigation. 

\begin{figure}
    \includegraphics[width=\columnwidth]{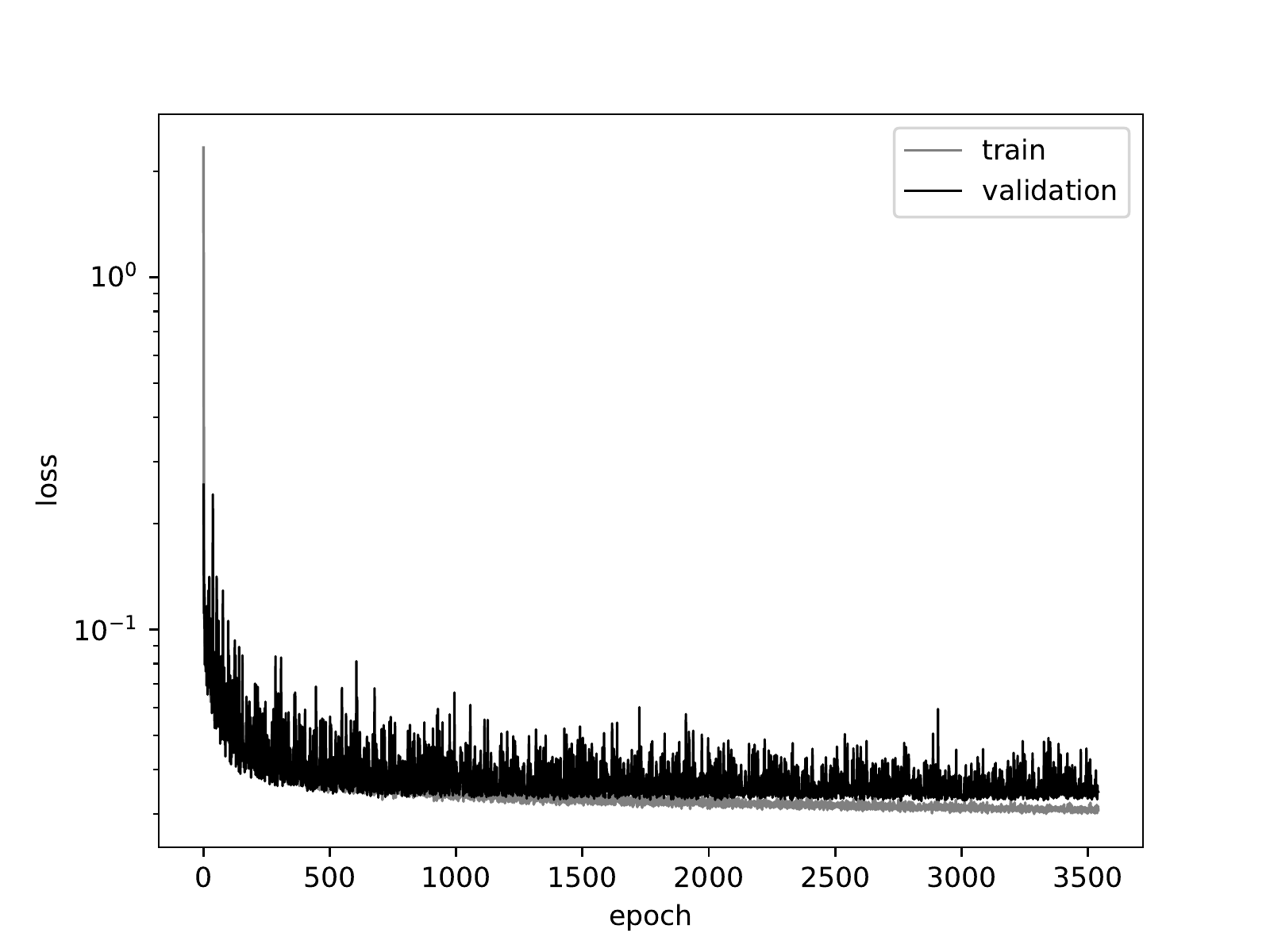}
    \caption{The above figure shows training and validation loss as number of epoch increases for the model predicting stellar mass. The gap between the training and validation loss is very small.}
    \label{fig:smloss_fig}
\end{figure}

\begin{figure}
    \includegraphics[width=\columnwidth]{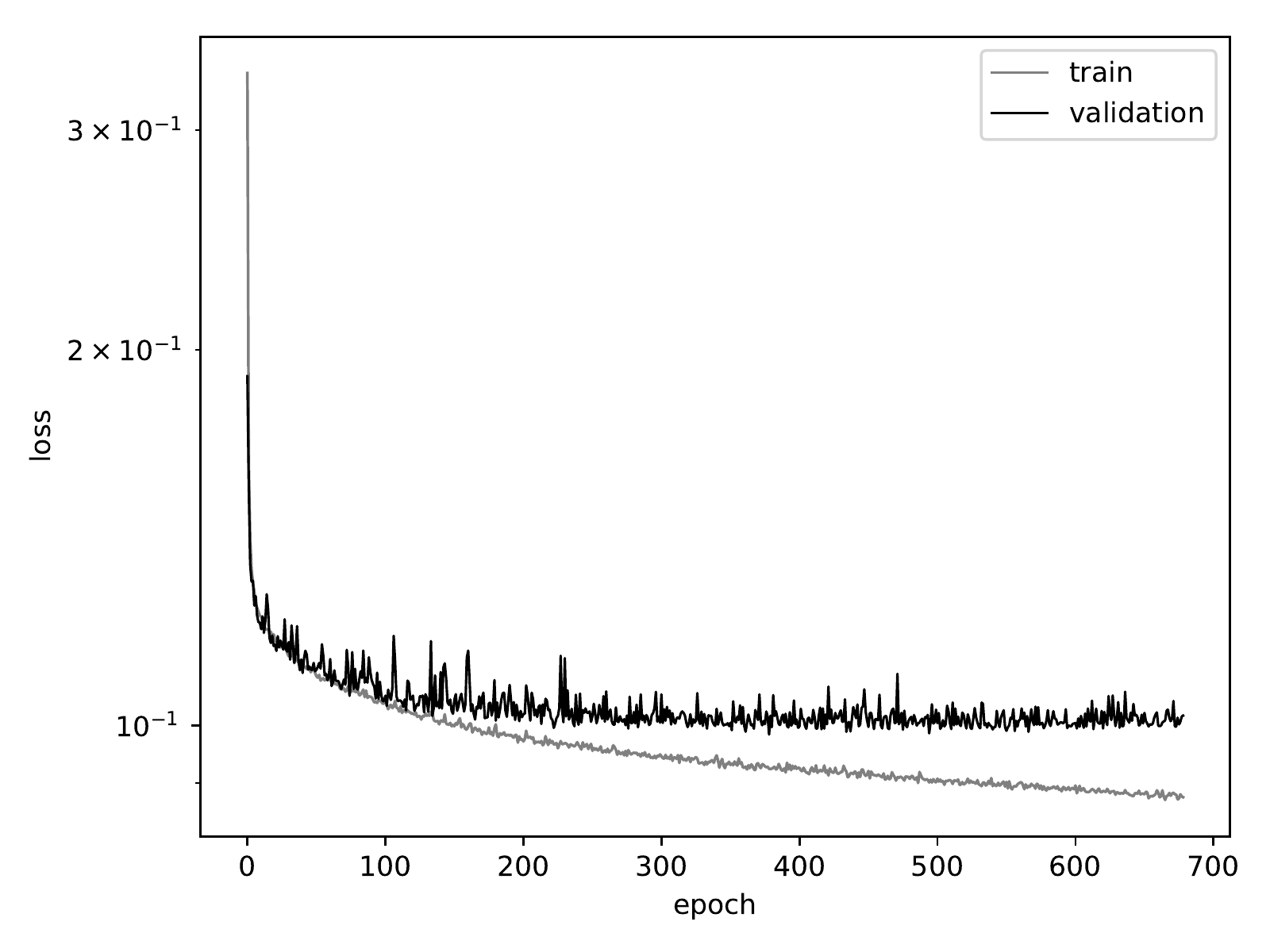}
    \caption{The above figure shows training and validation loss as number of epoch increases for the model predicting SFR. The gap between the training and validation loss was reduced using the early stopping method. It may be further reduced with increase in the number of data points available.}
    \label{fig:sfrloss_fig}
\end{figure}

\begin{figure}
    \includegraphics[width=\columnwidth]{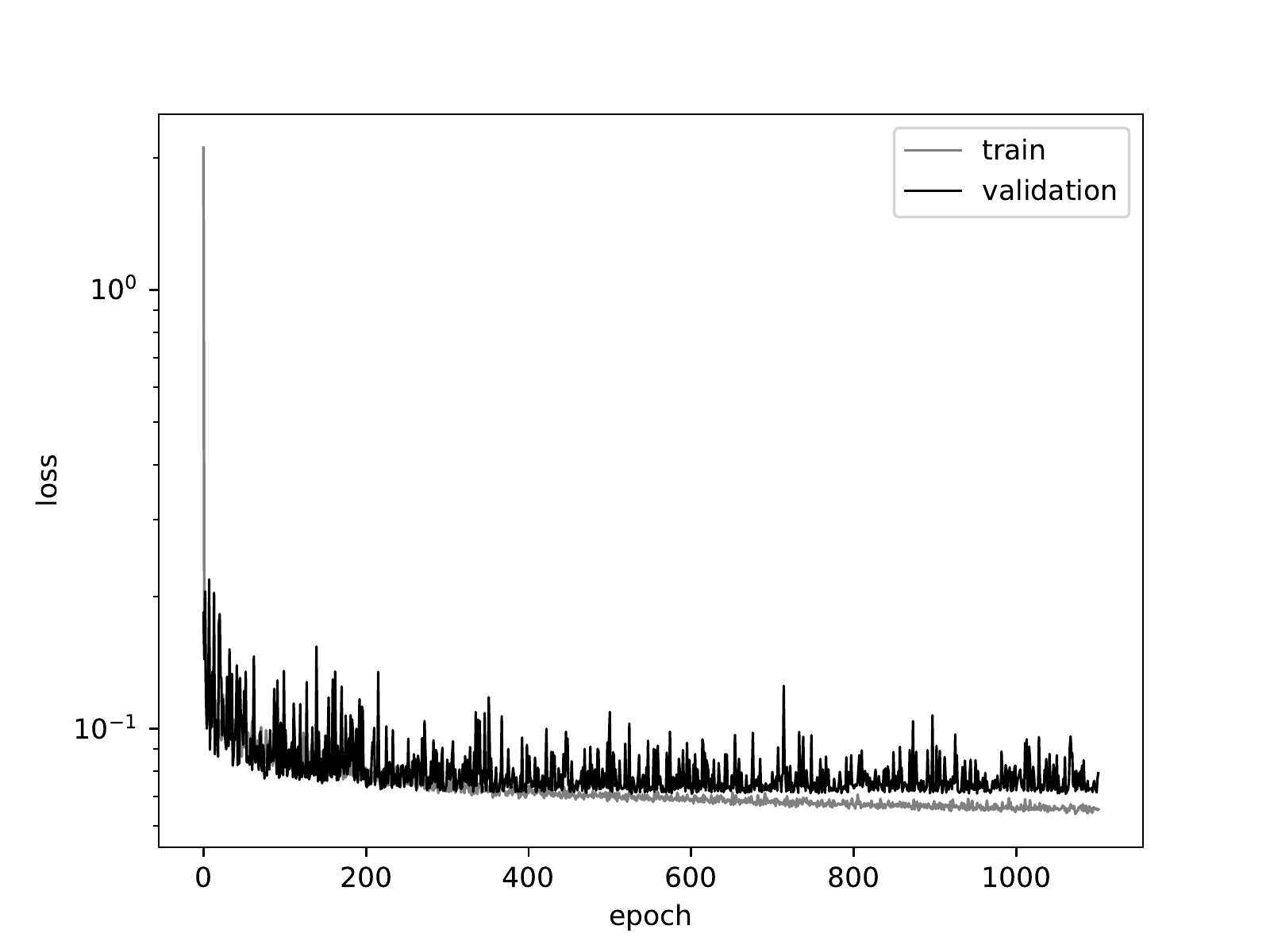}
    \caption{The above figure shows training and validation loss as number of epoch increases for the model predicting dust luminosity. The gap between the training and validation loss is very small.}
    \label{fig:dlloss_fig}
\end{figure}

Figs.~\ref{fig:smloss_fig}, \ref{fig:sfrloss_fig} and \ref{fig:dlloss_fig} show the loss of each of the models with increase in iteration. Note that for dust luminosity and stellar mass, the validation error closely follows the training error. Whereas, for SFR, the validation error is still slightly higher than the training error. This is after we used the early stopping technique to prevent the model from overfitting the training data (This gap was much larger without using early stopping). While this method helped, training on more data would help reduce this error further.
 
The deep learning technique outlined in this paper takes 3 to 30 minutes to train depending on the free parameter being modelled. Once we have a model, the time taken to predict the free parameters for the test data is negligible. To estimate the three star-formation parameters using the \magphys{} code for 10,000 galaxies would take $\sim100,000$ minutes (about 10 minutes per galaxy) and to predict the same number of galaxies using a deep learning model is $\sim30$ minutes with essentially all the time being taken up by training. This represents a huge savings in time, with potentially larger savings for samples from future large area imaging surveys. Further, this model can be modified to also give the confidence level of each prediction. Those galaxies whose free parameters the model predicts with low confidence can be investigated and then rerun with the standard stellar population technique. The outcome of this can then  be further investigated and changes incorporated in the deep learning model. Thus the model can get enriched as it encounters more and more data which enables it to capture more information and patterns contained in the data.

\section{Conclusions} \label{conclusion}

We have used deep neural networks to predict the free parameters - stellar mass, star formation rate and dust luminosity based on the multiband flux measurements from the GAMA survey. In our galaxy sample, each of these parameters spans three orders of magnitude. Over this large span, almost all the scatter is random in nature with little systematic. We caution, however, that our deep learning is only learning to emulate the \magphys{} model and not the galaxies in the real Universe. If \magphys{} models are incorrect or incomplete representations of galaxies in some region of the parameter space, our deep learning models will also learn these deficiencies. A positive aspect of this limitation, is that as stellar populations synthesis models improve in the future, deep learning can learn from these new models and improve its performance. 

Our approach reduces the time taken to derive the star-formation parameters drastically. These models can further be enhanced to give a confidence value with each prediction  and to incorporate the error associated with each input measurement. In this work, we used the spectroscopic redshift as an input parameter. In upcoming large area galaxy surveys with telescopes like the Large Synoptic Survey Telescope (LSST) it should become possible to  determine both the photometric redshift and the star-formation properties jointly using a deep learning approach. We look forward to the exciting times that lie ahead.

\section*{Acknowledgements}

SS thanks National Centre for Radio Astrophysics for hosting her during the time this work was completed.
GAMA is a joint European-Australasian project based around a spectroscopic campaign using the Anglo-Australian Telescope. The GAMA input catalogue is based on data taken from the Sloan Digital Sky Survey and the UKIRT Infrared Deep Sky Survey. Complementary imaging of the GAMA regions is being obtained by a number of independent survey programmes including GALEX MIS, VST KiDS, VISTA VIKING, WISE, Herschel-ATLAS, GMRT and ASKAP providing UV to radio coverage. GAMA is funded by the STFC (UK), the ARC (Australia), the AAO, and the participating institutions. The GAMA website is http://www.gama-survey.org/ .





\bibliographystyle{mnras}
\bibliography{bibliography.bib}

\begin{thebibliography}{}
\makeatletter
\relax
\def\mn@urlcharsother{\let\do\@makeother \do\$\do\&\do\#\do\^\do\_\do\%\do\~}
\def\mn@doi{\begingroup\mn@urlcharsother \@ifnextchar [ {\mn@doi@}
  {\mn@doi@[]}}
\def\mn@doi@[#1]#2{\def\@tempa{#1}\ifx\@tempa\@empty \href
  {http://dx.doi.org/#2} {doi:#2}\else \href {http://dx.doi.org/#2} {#1}\fi
  \endgroup}
\def\mn@eprint#1#2{\mn@eprint@#1:#2::\@nil}
\def\mn@eprint@arXiv#1{\href {http://arxiv.org/abs/#1} {{\tt arXiv:#1}}}
\def\mn@eprint@dblp#1{\href {http://dblp.uni-trier.de/rec/bibtex/#1.xml}
  {dblp:#1}}
\def\mn@eprint@#1:#2:#3:#4\@nil{\def\@tempa {#1}\def\@tempb {#2}\def\@tempc
  {#3}\ifx \@tempc \@empty \let \@tempc \@tempb \let \@tempb \@tempa \fi \ifx
  \@tempb \@empty \def\@tempb {arXiv}\fi \@ifundefined
  {mn@eprint@\@tempb}{\@tempb:\@tempc}{\expandafter \expandafter \csname
  mn@eprint@\@tempb\endcsname \expandafter{\@tempc}}}

\bibitem[\protect\citeauthoryear{{Abraham}, {Philip}, {Kembhavi}, {Wadadekar}
  \& {Sinha}}{{Abraham} et~al.}{2012}]{Abraham2012}
{Abraham} S.,  {Philip} N.~S.,  {Kembhavi} A.,  {Wadadekar} Y.~G.,   {Sinha}
  R.,  2012, \mn@doi [\mnras] {10.1111/j.1365-2966.2011.19674.x}, \href
  {https://ui.adsabs.harvard.edu/abs/2012MNRAS.419...80A} {419, 80}

\bibitem[\protect\citeauthoryear{{Bait}, {Barway}  \& {Wadadekar}}{{Bait}
  et~al.}{2017}]{Bait17}
{Bait} O.,  {Barway} S.,   {Wadadekar} Y.,  2017, \mn@doi [\mnras]
  {10.1093/mnras/stx1688}, \href
  {https://ui.adsabs.harvard.edu/abs/2017MNRAS.471.2687B} {471, 2687}

\bibitem[\protect\citeauthoryear{{Ball}, {Brunner}, {Myers}  \&
  {Tcheng}}{{Ball} et~al.}{2006}]{Ball2006}
{Ball} N.~M.,  {Brunner} R.~J.,  {Myers} A.~D.,   {Tcheng} D.,  2006, \mn@doi
  [\apj] {10.1086/507440}, \href
  {https://ui.adsabs.harvard.edu/abs/2006ApJ...650..497B} {650, 497}

\bibitem[\protect\citeauthoryear{{Ball}, {Brunner}, {Myers}, {Strand},
  {Alberts}  \& {Tcheng}}{{Ball} et~al.}{2008}]{Ball2008}
{Ball} N.~M.,  {Brunner} R.~J.,  {Myers} A.~D.,  {Strand} N.~E.,  {Alberts}
  S.~L.,   {Tcheng} D.,  2008, \mn@doi [\apj] {10.1086/589646}, \href
  {https://ui.adsabs.harvard.edu/abs/2008ApJ...683...12B} {683, 12}

\bibitem[\protect\citeauthoryear{{Banerji} et~al.,}{{Banerji}
  et~al.}{2010}]{Banerji2010}
{Banerji} M.,  et~al., 2010, \mn@doi [\mnras]
  {10.1111/j.1365-2966.2010.16713.x}, \href
  {https://ui.adsabs.harvard.edu/abs/2010MNRAS.406..342B} {406, 342}

\bibitem[\protect\citeauthoryear{{Barchi} et~al.,}{{Barchi}
  et~al.}{2019}]{Barchi2019}
{Barchi} P.~H.,  et~al., 2019, arXiv e-prints, \href
  {https://ui.adsabs.harvard.edu/abs/2019arXiv190107047B} {p. arXiv:1901.07047}

\bibitem[\protect\citeauthoryear{{Baron}}{{Baron}}{2019}]{Baron2019}
{Baron} D.,  2019, arXiv e-prints, \href
  {https://ui.adsabs.harvard.edu/abs/2019arXiv190407248B} {p. arXiv:1904.07248}

\bibitem[\protect\citeauthoryear{{Berta} et~al.,}{{Berta}
  et~al.}{2013}]{Berta13}
{Berta} S.,  et~al., 2013, \mn@doi [\aap] {10.1051/0004-6361/201220859}, \href
  {https://ui.adsabs.harvard.edu/abs/2013A&A...551A.100B} {551, A100}

\bibitem[\protect\citeauthoryear{{Brescia}, {Cavuoti}  \& {Longo}}{{Brescia}
  et~al.}{2015}]{Brescia2015}
{Brescia} M.,  {Cavuoti} S.,   {Longo} G.,  2015, \mn@doi [\mnras]
  {10.1093/mnras/stv854}, \href
  {https://ui.adsabs.harvard.edu/abs/2015MNRAS.450.3893B} {450, 3893}

\bibitem[\protect\citeauthoryear{{Bruzual} \& {Charlot}}{{Bruzual} \&
  {Charlot}}{2003}]{Bruzual2003}
{Bruzual} G.,  {Charlot} S.,  2003, \mn@doi [\mnras]
  {10.1046/j.1365-8711.2003.06897.x}, \href
  {https://ui.adsabs.harvard.edu/abs/2003MNRAS.344.1000B} {344, 1000}

\bibitem[\protect\citeauthoryear{{Chang}, {van der Wel}, {da Cunha}  \&
  {Rix}}{{Chang} et~al.}{2015}]{Chang15}
{Chang} Y.-Y.,  {van der Wel} A.,  {da Cunha} E.,   {Rix} H.-W.,  2015, \mn@doi
  [\apjs] {10.1088/0067-0049/219/1/8}, \href
  {https://ui.adsabs.harvard.edu/abs/2015ApJS..219....8C} {219, 8}

\bibitem[\protect\citeauthoryear{{Charlot} \& {Fall}}{{Charlot} \&
  {Fall}}{2000}]{Charlot2000}
{Charlot} S.,  {Fall} S.~M.,  2000, \mn@doi [\apj] {10.1086/309250}, \href
  {https://ui.adsabs.harvard.edu/abs/2000ApJ...539..718C} {539, 718}

\bibitem[\protect\citeauthoryear{{Conroy}}{{Conroy}}{2013}]{Conroy2013}
{Conroy} C.,  2013, \mn@doi [\araa] {10.1146/annurev-astro-082812-141017},
  \href {https://ui.adsabs.harvard.edu/abs/2013ARA&A..51..393C} {51, 393}

\bibitem[\protect\citeauthoryear{{D'Isanto}, {Cavuoti}, {Brescia}, {Donalek},
  {Longo}, {Riccio}  \& {Djorgovski}}{{D'Isanto} et~al.}{2016}]{D'Isanto2016}
{D'Isanto} A.,  {Cavuoti} S.,  {Brescia} M.,  {Donalek} C.,  {Longo} G.,
  {Riccio} G.,   {Djorgovski} S.~G.,  2016, \mn@doi [\mnras]
  {10.1093/mnras/stw157}, \href
  {https://ui.adsabs.harvard.edu/abs/2016MNRAS.457.3119D} {457, 3119}

\bibitem[\protect\citeauthoryear{{Delli Veneri}, {Cavuoti}, {Brescia}, {Longo}
  \& {Riccio}}{{Delli Veneri} et~al.}{2019}]{delli2019}
{Delli Veneri} M.,  {Cavuoti} S.,  {Brescia} M.,  {Longo} G.,   {Riccio} G.,
  2019, \mn@doi [\mnras] {10.1093/mnras/stz856}, \href
  {https://ui.adsabs.harvard.edu/abs/2019MNRAS.486.1377D} {486, 1377}

\bibitem[\protect\citeauthoryear{{Driver} et~al.,}{{Driver}
  et~al.}{2011}]{Driver2011}
{Driver} S.~P.,  et~al., 2011, \mn@doi [\mnras]
  {10.1111/j.1365-2966.2010.18188.x}, \href
  {https://ui.adsabs.harvard.edu/abs/2011MNRAS.413..971D} {413, 971}

\bibitem[\protect\citeauthoryear{{Driver} et~al.,}{{Driver}
  et~al.}{2016}]{Driver2016}
{Driver} S.~P.,  et~al., 2016, \mn@doi [\mnras] {10.1093/mnras/stv2505}, \href
  {https://ui.adsabs.harvard.edu/abs/2016MNRAS.455.3911D} {455, 3911}

\bibitem[\protect\citeauthoryear{{Driver} et~al.,}{{Driver}
  et~al.}{2018}]{Driver18}
{Driver} S.~P.,  et~al., 2018, \mn@doi [\mnras] {10.1093/mnras/stx2728}, \href
  {https://ui.adsabs.harvard.edu/abs/2018MNRAS.475.2891D} {475, 2891}

\bibitem[\protect\citeauthoryear{{Forbes}, {Krumholz}  \& {Speagle}}{{Forbes}
  et~al.}{2019}]{Forbes2019}
{Forbes} J.~C.,  {Krumholz} M.~R.,   {Speagle} J.~S.,  2019, \mn@doi [\mnras]
  {10.1093/mnras/stz1473}, \href
  {https://ui.adsabs.harvard.edu/abs/2019MNRAS.487.3581F} {487, 3581}

\bibitem[\protect\citeauthoryear{Glorot, Bordes  \& Bengio}{Glorot
  et~al.}{2011}]{glorot2011deep}
Glorot X.,  Bordes A.,   Bengio Y.,  2011, in Proceedings of the fourteenth
  international conference on artificial intelligence and statistics. pp
  315--323

\bibitem[\protect\citeauthoryear{Goodfellow, Bengio  \& Courville}{Goodfellow
  et~al.}{2016}]{Goodfellow2016}
Goodfellow I.,  Bengio Y.,   Courville A.,  2016, Deep Learning.
MIT Press

\bibitem[\protect\citeauthoryear{{Guill{\'e}n}, {Bueno}, {Carceller},
  {Mart{\'\i}nez-Vel{\'a}zquez}, {Rubio}, {Todero Peixoto}  \&
  {Sanchez-Lucas}}{{Guill{\'e}n} et~al.}{2019}]{Guill2019}
{Guill{\'e}n} A.,  {Bueno} A.,  {Carceller} J.~M.,
  {Mart{\'\i}nez-Vel{\'a}zquez} J.~C.,  {Rubio} G.,  {Todero Peixoto} C.~J.,
  {Sanchez-Lucas} P.,  2019, \mn@doi [Astroparticle Physics]
  {10.1016/j.astropartphys.2019.03.001}, \href
  {https://ui.adsabs.harvard.edu/abs/2019APh...111...12G} {111, 12}

\bibitem[\protect\citeauthoryear{{Harp} et~al.,}{{Harp}
  et~al.}{2019}]{Harp2019}
{Harp} G.~R.,  et~al., 2019, arXiv e-prints, \href
  {https://ui.adsabs.harvard.edu/abs/2019arXiv190202426H} {p. arXiv:1902.02426}

\bibitem[\protect\citeauthoryear{{Hemmati} et~al.,}{{Hemmati}
  et~al.}{2019}]{hemmati2019}
{Hemmati} S.,  et~al., 2019, arXiv e-prints, \href
  {https://ui.adsabs.harvard.edu/abs/2019arXiv190510379H} {p. arXiv:1905.10379}

\bibitem[\protect\citeauthoryear{Hinton, Osindero  \& Teh}{Hinton
  et~al.}{2006}]{hinton2006fast}
Hinton G.~E.,  Osindero S.,   Teh Y.-W.,  2006, Neural computation, 18, 1527

\bibitem[\protect\citeauthoryear{{Huertas-Company}, {Rouan}, {Tasca}, {Soucail}
   \& {Le F{\`e}vre}}{{Huertas-Company} et~al.}{2008}]{Huertas-Company08}
{Huertas-Company} M.,  {Rouan} D.,  {Tasca} L.,  {Soucail} G.,   {Le F{\`e}vre}
  O.,  2008, \mn@doi [\aap] {10.1051/0004-6361:20078625}, \href
  {https://ui.adsabs.harvard.edu/abs/2008A&A...478..971H} {478, 971}

\bibitem[\protect\citeauthoryear{{Huertas-Company}, {Aguerri}, {Bernardi},
  {Mei}  \& {S{\'a}nchez Almeida}}{{Huertas-Company}
  et~al.}{2011}]{Huertas-Company11}
{Huertas-Company} M.,  {Aguerri} J.~A.~L.,  {Bernardi} M.,  {Mei} S.,
  {S{\'a}nchez Almeida} J.,  2011, \mn@doi [\aap]
  {10.1051/0004-6361/201015735}, \href
  {https://ui.adsabs.harvard.edu/abs/2011A&A...525A.157H} {525, A157}

\bibitem[\protect\citeauthoryear{Kingma \& Ba}{Kingma \&
  Ba}{2014}]{kingma2014adam}
Kingma D.~P.,  Ba J.,  2014, arXiv preprint arXiv:1412.6980

\bibitem[\protect\citeauthoryear{LeCun, Bengio  \& Hinton}{LeCun
  et~al.}{2015}]{lecun2015deep}
LeCun Y.,  Bengio Y.,   Hinton G.,  2015, nature, 521, 436

\bibitem[\protect\citeauthoryear{{Lochner}, {McEwen}, {Peiris}, {Lahav}  \&
  {Winter}}{{Lochner} et~al.}{2016}]{Lochner2016}
{Lochner} M.,  {McEwen} J.~D.,  {Peiris} H.~V.,  {Lahav} O.,   {Winter} M.~K.,
  2016, \mn@doi [\apjs] {10.3847/0067-0049/225/2/31}, \href
  {https://ui.adsabs.harvard.edu/abs/2016ApJS..225...31L} {225, 31}

\bibitem[\protect\citeauthoryear{{Lovell}, {Acquaviva}, {Thomas}, {Iyer},
  {Gawiser}  \& {Wilkins}}{{Lovell} et~al.}{2019}]{Lovell2019}
{Lovell} C.~C.,  {Acquaviva} V.,  {Thomas} P.~A.,  {Iyer} K.~G.,  {Gawiser} E.,
    {Wilkins} S.~M.,  2019, arXiv e-prints, \href
  {https://ui.adsabs.harvard.edu/abs/2019arXiv190310457L} {p. arXiv:1903.10457}

\bibitem[\protect\citeauthoryear{{Lukic}, {Br{\"u}ggen}, {Banfield}, {Wong},
  {Rudnick}, {Norris}  \& {Simmons}}{{Lukic} et~al.}{2018}]{Lukic2018}
{Lukic} V.,  {Br{\"u}ggen} M.,  {Banfield} J.~K.,  {Wong} O.~I.,  {Rudnick} L.,
   {Norris} R.~P.,   {Simmons} B.,  2018, \mn@doi [\mnras]
  {10.1093/mnras/sty163}, \href
  {https://ui.adsabs.harvard.edu/abs/2018MNRAS.476..246L} {476, 246}

\bibitem[\protect\citeauthoryear{McKinney}{McKinney}{2010}]{mckinney-proc-scipy-2010}
McKinney W.,  2010, in van~der Walt S.,  Millman J.,  eds, Proceedings of the
  9th Python in Science Conference. pp 51 -- 56

\bibitem[\protect\citeauthoryear{Mitchell}{Mitchell}{1997}]{Mitchell:1997:ML:541177}
Mitchell T.~M.,  1997, Machine Learning, 1 edn.
McGraw-Hill, Inc., New York, NY, USA

\bibitem[\protect\citeauthoryear{Nair \& Hinton}{Nair \&
  Hinton}{2010}]{Nair2010RectifiedLU}
Nair V.,  Hinton G.~E.,  2010, in ICML.

\bibitem[\protect\citeauthoryear{{Pearson}, {Li}  \& {Dye}}{{Pearson}
  et~al.}{2019}]{Pearson2019}
{Pearson} J.,  {Li} N.,   {Dye} S.,  2019, \mn@doi [\mnras]
  {10.1093/mnras/stz1750}, \href
  {https://ui.adsabs.harvard.edu/abs/2019MNRAS.tmp.1699P} {p.~1699}

\bibitem[\protect\citeauthoryear{Pedregosa et~al.,}{Pedregosa
  et~al.}{2011}]{Pedregosa:2011:SML:1953048.2078195}
Pedregosa F.,  et~al., 2011, J. Mach. Learn. Res., 12, 2825

\bibitem[\protect\citeauthoryear{{Philip}, {Wadadekar}, {Kembhavi}  \&
  {Joseph}}{{Philip} et~al.}{2002}]{Philip2002}
{Philip} N.~S.,  {Wadadekar} Y.,  {Kembhavi} A.,   {Joseph} K.~B.,  2002,
  \mn@doi [\aap] {10.1051/0004-6361:20020219}, \href
  {https://ui.adsabs.harvard.edu/abs/2002A&A...385.1119P} {385, 1119}

\bibitem[\protect\citeauthoryear{{Schaye} et~al.,}{{Schaye}
  et~al.}{2015}]{Schaye2015}
{Schaye} J.,  et~al., 2015, \mn@doi [\mnras] {10.1093/mnras/stu2058}, \href
  {https://ui.adsabs.harvard.edu/abs/2015MNRAS.446..521S} {446, 521}

\bibitem[\protect\citeauthoryear{{Stensbo-Smidt}, {Gieseke}, {Igel}, {Zirm}  \&
  {Steenstrup Pedersen}}{{Stensbo-Smidt} et~al.}{2017}]{Stensbo-Smidt2017}
{Stensbo-Smidt} K.,  {Gieseke} F.,  {Igel} C.,  {Zirm} A.,   {Steenstrup
  Pedersen} K.,  2017, \mn@doi [\mnras] {10.1093/mnras/stw2476}, \href
  {https://ui.adsabs.harvard.edu/abs/2017MNRAS.464.2577S} {464, 2577}

\bibitem[\protect\citeauthoryear{Sutskever, Hinton  \& Krizhevsky}{Sutskever
  et~al.}{2012}]{sutskever2012imagenet}
Sutskever I.,  Hinton G.~E.,   Krizhevsky A.,  2012, Advances in neural
  information processing systems, pp 1097--1105

\bibitem[\protect\citeauthoryear{{Viaene} et~al.,}{{Viaene}
  et~al.}{2014}]{Viaene14}
{Viaene} S.,  et~al., 2014, \mn@doi [\aap] {10.1051/0004-6361/201423534}, \href
  {https://ui.adsabs.harvard.edu/abs/2014A&A...567A..71V} {567, A71}

\bibitem[\protect\citeauthoryear{{Vogelsberger} et~al.,}{{Vogelsberger}
  et~al.}{2014}]{Vogelsberger2014}
{Vogelsberger} M.,  et~al., 2014, \mn@doi [\nat] {10.1038/nature13316}, \href
  {https://ui.adsabs.harvard.edu/abs/2014Natur.509..177V} {509, 177}

\bibitem[\protect\citeauthoryear{{Wadadekar}}{{Wadadekar}}{2005}]{Wadadekar2005}
{Wadadekar} Y.,  2005, \mn@doi [\pasp] {10.1086/427710}, \href
  {https://ui.adsabs.harvard.edu/abs/2005PASP..117...79W} {117, 79}

\bibitem[\protect\citeauthoryear{Witten, Frank, Hall  \& Pal}{Witten
  et~al.}{2016}]{witten2016data}
Witten I.~H.,  Frank E.,  Hall M.~A.,   Pal C.~J.,  2016, Data Mining:
  Practical machine learning tools and techniques.
Morgan Kaufmann

\bibitem[\protect\citeauthoryear{Zeiler et~al.,}{Zeiler
  et~al.}{2013}]{zeiler2013rectified}
Zeiler M.~D.,  et~al., 2013, in 2013 IEEE International Conference on
  Acoustics, Speech and Signal Processing. pp 3517--3521

\bibitem[\protect\citeauthoryear{{Zhang} \& {Zhao}}{{Zhang} \&
  {Zhao}}{2006}]{Zhang2006}
{Zhang} Y.,  {Zhao} Y.,  2006, arXiv e-prints, \href
  {https://ui.adsabs.harvard.edu/abs/2006astro.ph.12727Z} {pp
  astro--ph/0612727}

\bibitem[\protect\citeauthoryear{{da Cunha}, {Charlot}  \& {Elbaz}}{{da Cunha}
  et~al.}{2008}]{daCunha2008}
{da Cunha} E.,  {Charlot} S.,   {Elbaz} D.,  2008, \mn@doi [\mnras]
  {10.1111/j.1365-2966.2008.13535.x}, \href
  {https://ui.adsabs.harvard.edu/abs/2008MNRAS.388.1595D} {388, 1595}

\bibitem[\protect\citeauthoryear{{da Cunha}, {Charmandaris},
  {D{\'\i}az-Santos}, {Armus}, {Marshall}  \& {Elbaz}}{{da Cunha}
  et~al.}{2010}]{daCunha10}
{da Cunha} E.,  {Charmandaris} V.,  {D{\'\i}az-Santos} T.,  {Armus} L.,
  {Marshall} J.~A.,   {Elbaz} D.,  2010, \mn@doi [\aap]
  {10.1051/0004-6361/201014498}, \href
  {https://ui.adsabs.harvard.edu/abs/2010A&A...523A..78D} {523, A78}

\makeatother
\end{thebibliography}


\bsp	
\label{lastpage}

\end{document}